\documentclass[prd,showpcs,amsmath,amssymb,nofootinbib,longbibliography,twocolumn,showpacs,notitlepage
]{revtex4-1}
\usepackage{multirow}
\usepackage{epsfig}
\usepackage{amsmath}
\usepackage{bm}
\usepackage{times}
\usepackage{graphicx}
\usepackage{color} 
\usepackage{slashed}
\usepackage{graphicx}
\usepackage{amsmath}
\usepackage[latin1]{inputenc}
\usepackage{hyperref}
\usepackage{soul}
\usepackage{multirow}
\usepackage{epsfig}
\usepackage{amsmath}
\usepackage{bm}
\usepackage{times}
\usepackage{graphicx}
\usepackage{color}
\usepackage{slashed}
\usepackage{graphicx}
\usepackage{amsmath}
\usepackage{relsize} 
\usepackage[latin1]{inputenc}

\newcommand{\overbar}[1]{\mkern 1.5mu\overline{\mkern-1.5mu#1\mkern-1.5mu}\mkern 1.5mu}

\def\bea{\begin{eqnarray}}
\def\eea{\end{eqnarray}}
\def\bean{\begin{eqnarray*}}
\def\eean{\end{eqnarray*}} 
\def\nn{\nonumber}
\def\beaal{\begin{align}}
\def\eeaal{\end{align}}

\begin{document} 
 
\title{Baryonic and Leptonic GeV Dark Matter}

\author{Bartosz~Fornal}
\affiliation{\vspace{1mm}Department of Physics and Astronomy, University of Utah, Salt Lake City, UT 84112, USA\vspace{1mm}}
\author{Alec~Hewitt}
\affiliation{\vspace{1mm}Department of Physics and Astronomy, University of Utah, Salt Lake City, UT 84112, USA\vspace{1mm}}
\author{Yue~Zhao}
\affiliation{\vspace{1mm}Department of Physics and Astronomy, University of Utah, Salt Lake City, UT 84112, USA\vspace{1mm}}
\date{\today}
\begin{abstract}
\vspace{2mm}
We perform a systematic analysis of  models with GeV-scale dark matter coupled to  baryons and leptons.  Such theories provide
 a natural framework to explain the matter-antimatter asymmetry of the universe. We find that only a few baryonic dark matter models are free from tree-level proton decay without explicitly imposing baryon number conservation.
 We enumerate those cases and provide a brief overview of their phenomenology.  We then focus on a leptonic dark matter model  for a more detailed discussion of the baryon asymmetry generation via leptogenesis, the  symmetry restoration in the dark sector and the expected dark matter annihilation signals in indirect detection experiments.\vspace{6mm}
\end{abstract}

\maketitle

\vspace{6mm}

\noindent
\section{Introduction}

The evidence for  dark matter in the universe is indisputable. Not only its existence, but also its  distribution and abundance are  precisely determined from various measurements, including
 galaxy rotation curves \cite{Vera}, cosmic microwave background  \cite{deBernardis:2000sbo}, gravitational lensing \cite{Gavazzi:2007vw}, etc. 
Nevertheless, 
the dark matter mass and its  interactions with Standard Model particles remain a mystery. It is not even known if dark matter consists  of elementary particles or macroscopic objects. The mass of an elementary dark matter particle can be anywhere between  $\sim 10^{-31} \ {\rm GeV}$ (fuzzy dark matter) \cite{Press:1989id,Hui:2016ltb} and $\sim 10^{19} \ {\rm GeV}$ (WIMPzillas) \cite{Kolb:1998ki,Meissner:2018cay} (despite the unitarity bound \cite{Griest:1989wd}), while the mass of macroscopic  dark matter objects ranges from $\sim 10^{17}  \ {\rm GeV}$ (dark quark nuggets) \cite{Bai:2018dxf} to $\sim 10^{59} \ {\rm GeV}$ (primordial black holes) \cite{Carr:2009jm,Bird:2016dcv}. In most cases, the dark matter interactions with the known particles are small; from a theoretical perspective such interactions can even be absent. 

 Interestingly, the ratio of the abundances of dark matter and  ordinary matter is on the order of five. This suggests  that the two sectors may be related and, perhaps, share a common origin. This is precisely the idea behind theories of asymmetric dark matter \cite{Kaplan:2009ag,Zurek:2013wia}, in which an effective interaction  between dark matter and Standard Model particles  is established. 
The explanation of the matter-antimatter asymmetry of the universe in those theories relies on the asymmetries in the two sectors being generated simultaneously and the dark matter particles being at the GeV scale.

In this paper, we systematically analyze   scenarios in which  dark matter couples  to quarks and/or leptons. 
We study the possible effective operators describing such interactions up to dimension eight and analyze their particle model realizations. For a successful baryogenesis or leptogenesis, those operators must have a nonzero baryon number contribution carried by the  quarks or a nonzero lepton number from the leptons. 
The effective dark matter-Standard Model interactions require either scalar or vector mediators to be present at the particle level of each model. 
For dark  matter coupled to baryons \hspace{3mm} (baryonic dark matter) at least one mediator in each model is necessarily a color triplet. For dark matter coupled to leptons (leptonic dark matter)  the  mediators do not carry color. 

\vspace{8mm}

We demonstrate that for baryonic and baryoleptonic dark matter the possible mediators are:  the scalars $(3,1)_{-4/3}$, $(3,1)_{-1/3}$, $(3,1)_{2/3}$, or the vectors $(3,2)_{-5/6}$, $(3,2)_{1/6}$. Without imposing an additional symmetry, all of those particles, except for the scalar  $(3,1)_{2/3}$, can trigger tree-level proton decay \cite{Arnold:2012sd,Assad:2017iib}. Thus the mass of those mediators is elevated above $\sim 10^{16} \ {\rm GeV}$ by the stringent experimental constraints  on the proton lifetime, limiting their capabilities of explaining baryogenesis and their experimental probes.

Such large mediator masses can be avoided by fine-tuning some of the couplings to be small or by explicitly imposing baryon and/or lepton number conservation.  However, there is no  strong theoretical argument to expect baryon or lepton number to be exact symmetries of nature. In fact, both of them are already  violated within the Standard Model itself at  the non-perturbative level  by the electroweak sphalerons.
Guided by the requirement of proton stability  without assuming a fine-tuning of couplings or imposing an additional symmetry, we  focus on  baryonic dark matter models involving only the scalar mediator $(3,1)_{2/3}$, as well as on leptonic dark matter models. The latter  are naturally free from tree-level proton decay, since they  involve the color singlet scalar mediators $(1,1)_{1}$ or $(1,2)_{-1/2}$. 

We find that only some of the baryonic dark matter direct detection signatures considered in the literature are realized  in  models that are  free from proton decay. In particular, the dark matter-nucleon annihilation, which is possible in such models, necessarily involves a kaon in the final state.  This conclusion increases the importance  of the Deep Underground Neutrino Experiment (DUNE) \cite{Abi:2020wmh} in complementing the efforts of  Super-Kamiokande \cite{Fukuda:2002uc} in the search for baryonic dark matter. Regarding  leptonic dark matter models, based on a concrete example, we analyze their potential for generating the matter-antimatter asymmetry through leptogenesis and  highlight their  unique feature  of symmetry restoration in the dark sector, achieved when one of the  dark sector particles is unstable on long time scales and decays to dark matter. 
Such a rebirth of symmetric  dark matter  leads to 
the possibility of enhanced annihilation signals at present times, previously considered  in  the context of  heavy dark matter \cite{Cohen:2009fz,Bell:2010qt,Falkowski:2011xh} and  oscillating dark matter \cite{Buckley:2011ye,Cirelli:2011ac,Tulin:2012re,Okada:2012rm,Ibe:2019yra},
and results in  signatures that can be searched for by the Fermi Gamma-Ray Space Telescope \cite{Atwood:2009ez} and  the future e-ASTROGAM \cite{DeAngelis:2017gra}.

\section{The models}\label{s2}

In this section, we enumerate the possible particle physics models for the effective operators describing the  interactions of dark matter with  just the quarks (baryonic dark matter),  both quarks and leptons (baryoleptonic dark matter), and  with leptons only (leptonic dark matter), and briefly discuss their phenomenology.
We adopt  the four-component Dirac spinor notation for the fermion fields, indicating with a subscript $L$ or $R$ their left- or right-handed chirality. The conjugate fields are denoted by a bar symbol.
We focus on those operators which carry a nonzero baryon number contribution from quarks or a nonzero lepton number contribution from leptons, since only those theories provide promising asymmetric dark matter frameworks.  In particular, we do not consider operators of the type $q\bar{q} \chi\bar\chi$ or $l\bar{l} \chi\bar\chi$, which arise, e.g.,  in certain theories with gauge bosons coupled to baryon  or lepton number \cite{Duerr:2013dza,Schwaller:2013hqa}. 
In the following analysis,  $\chi$ and $\widetilde\chi$  are Dirac fermions, whereas $\phi$ and $\Phi$  are complex scalars. Both  $\chi$ and $\phi$  are Standard Model singlets.

\vspace{7mm}

\noindent
\begin{centerline}
{\bf{Baryonic dark matter}}
\end{centerline}

\begin{figure}[t!]
\includegraphics[trim={0.0cm 0.0cm 0 0},clip,width=4.8cm]{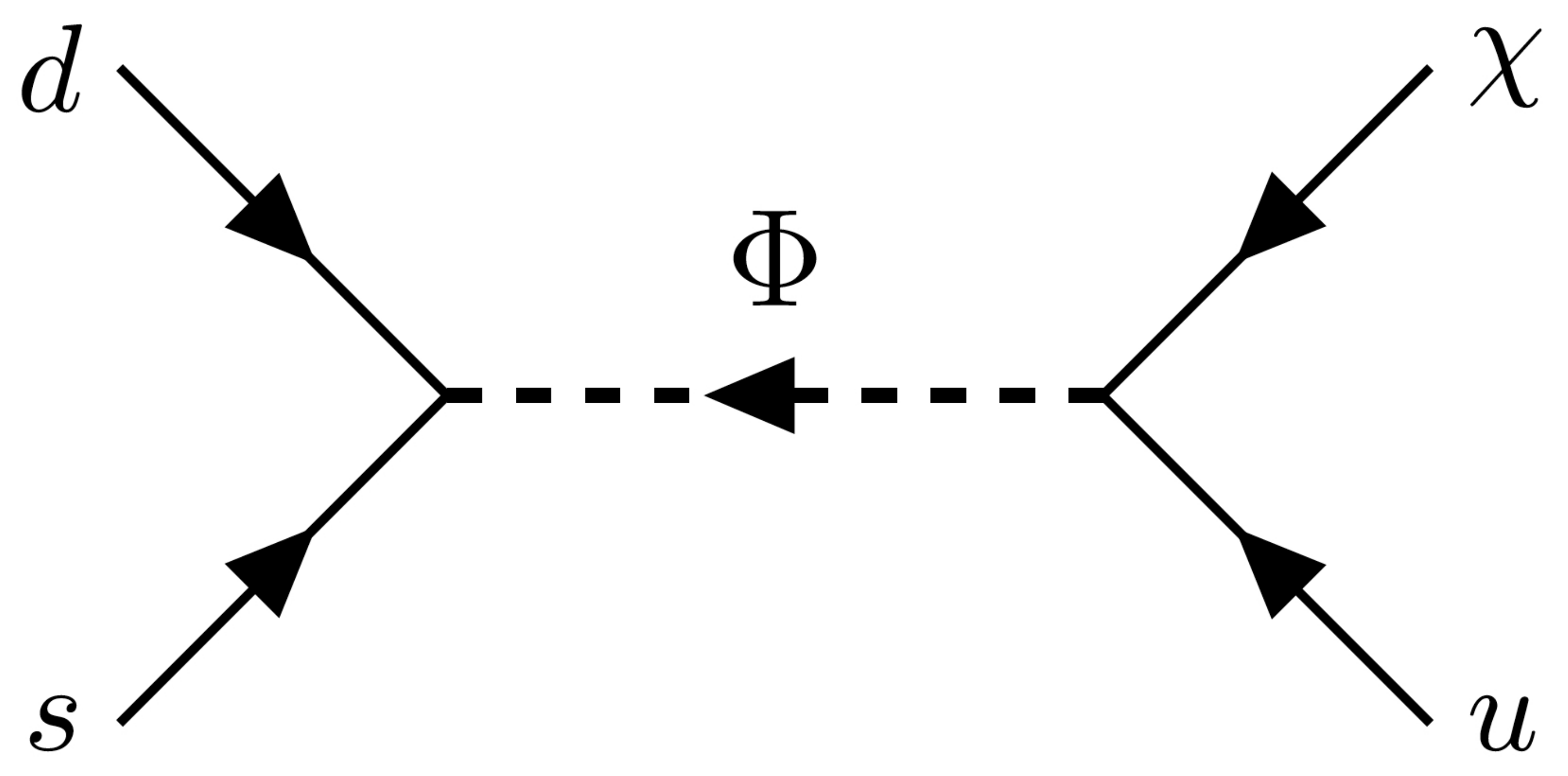}
\caption{{\small{Model 1A realization of the operator $qqq\chi$.\vspace{6mm}}  }}
\label{fig:1}
\end{figure}

The simplest effective operator  describing the interaction of dark matter with quarks is the dimension six $qqq\chi$, where $q$ stands for   $Q_L$, $d_R$ or $u_R$. The possible gauge-invariant realizations are $u_Rd_Rd_R\chi$ and $Q_LQ_Ld_R\chi$. There are three models one can write down for those operators, labeled as Models 1A--1C in Table \ref{tab33B}. They involve the mediators: scalar $(3,1)_{2/3}$, scalar $(3,1)_{-1/3}$, and  vector $(3,2)_{1/6}$, respectively.  Among those particles, only the scalar $(3,1)_{2/3}$ in Model 1A  does not give rise to tree-level proton decay, since the quantum numbers do not allow $\Phi$ to couple to a quark and a lepton \cite{Arnold:2012sd,Assad:2017iib}. This is in contrast to the other two cases, where  the scalar $(3,1)_{-1/3}$ can have the couplings  $\Phi\, u_R d_R$  and $\Phi^*u_R e_R$, whereas the vector  $(3,2)_{1/6}$  can couple via $X_\mu Q_L \gamma^\mu d_R$ and $X^\dagger_\mu L_L \gamma^\mu u_R$, both  leading to $p\to e^+ \pi^0$. For this reason, we focus on Model 1A below. An example of a diagram generating the operator  $qqq\chi$ is shown in Fig.\,{\ref{fig:1}}. 

\begin{figure}[t!]
\includegraphics[trim={0.0cm 0.0cm 0 -0.4cm},clip,width=5.7cm]{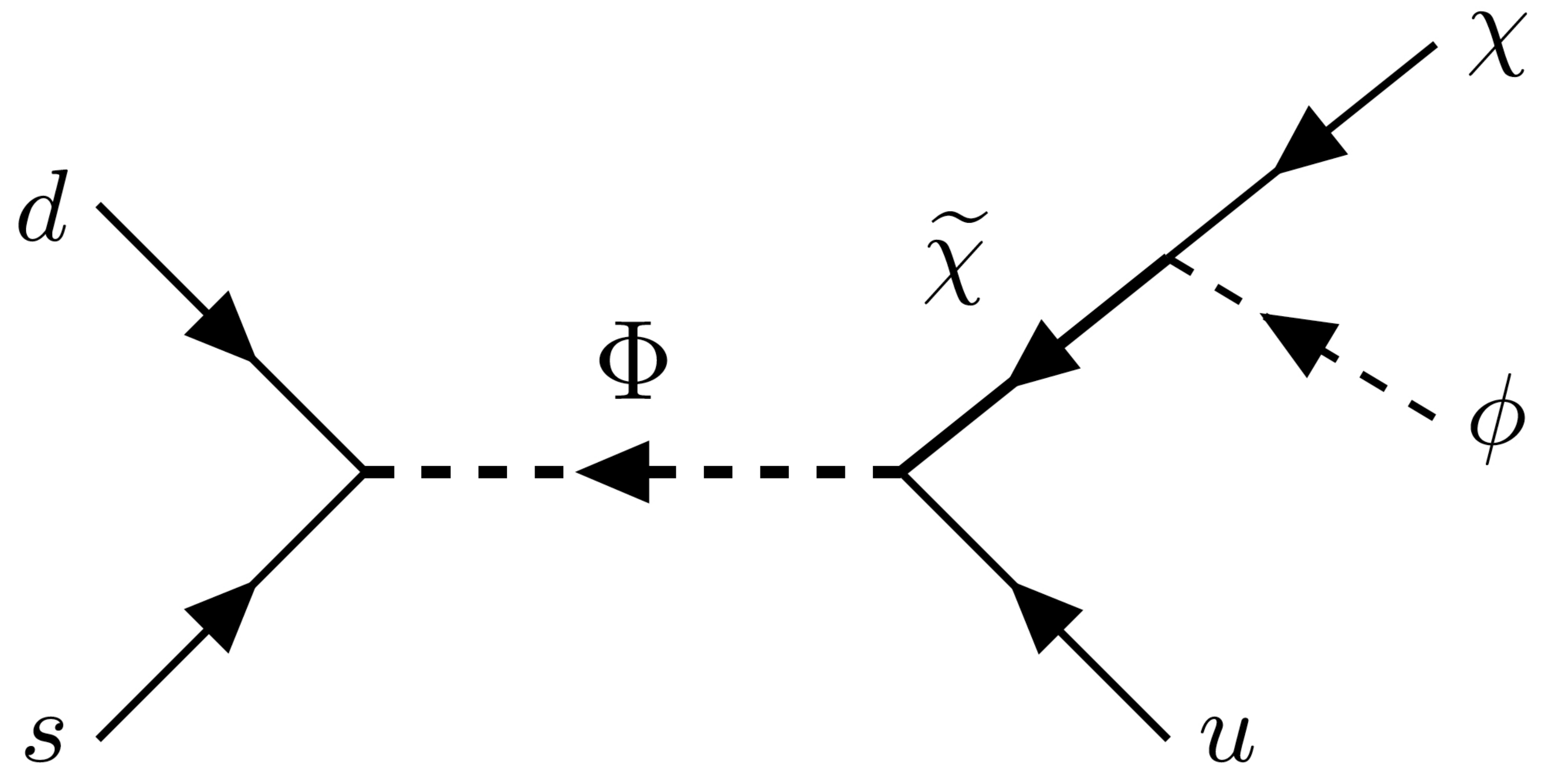}
\caption{{\small{Model 2A realization of the operator $qqq\chi\phi$.\vspace{2mm}}  }}
\label{fig:2}
\end{figure}

At dimension seven, the possible  effective interactions are  $qqq\chi\phi$ and $qqqH\chi$, where $H$ is the Higgs field. In this study, we do not consider operators involving the Higgs field and only focus on the ones in the first category.  There are six  particle models for the operator $qqq\chi\phi$,   denoted as Models 2A--2F in Table \ref{tab33B}. They involve  the same scalar and vector mediators as introduced for the operator $qqq\chi$. In addition, an intermediate fermion $\widetilde\chi$ is required (see, Fig.\,\ref{fig:2}). We discuss in more detail Model 2A, since, similarly to Model 1A, it does not suffer from tree-level proton decay. 

At dimension eight,  the operators are $qqq\chi\phi^2$, $qqqH\chi\phi$ and $qqqH^2\chi$. Again, we only consider the ones in the first category. Models for the operator $qqq\chi\phi^2$  can be constructed by introducing an additional scalar singlet field $\phi_e$, replacing $\phi$ with $\phi_e$ in Models 2A--2F,  and  adding the interaction $\phi_e^* \phi^2$. The particle  $\phi$ is then automatically stable without imposing specific relations between the masses. The particle $\chi$, on the other hand,  may be unstable if sufficiently heavy.

\begin{table}[t!] 
\begin{center}
\begingroup
\setlength{\tabcolsep}{6pt} 
\renewcommand{\arraystretch}{1.5} 
\begin{tabular}{  |c |c |c|c|} 
\hline
 \multicolumn{4}{|c|}{Baryonic dark matter}\\
\hline
\!\!Model\!\! & Interactions &  \multicolumn{2}{|c|}{Mediators} \\ 
\hline
\hline
\multicolumn{4}{|c|}{\raisebox{0pt}{${\boldsymbol{qqq\chi}}$}}\\[2pt]
\hline
1A &  \raisebox{0pt}{$\Phi \,d_R d_R$ , $\Phi^*\!u_R\chi$} & \multicolumn{2}{|c|}{$\Phi = (3,1)_{\frac23}$ \ \,}  \\[2pt]
\cline{1-4}
\raisebox{-5pt}{1B}  & \raisebox{0pt}{$\big(\Phi\, u_Rd_R$    {\rm or} \,$\Phi \,Q_LQ_L\big)$,} & \multicolumn{2}{|c|}{\raisebox{-5pt}{$\Phi = (3,1)_{-\frac13}$}}  \\[-10pt]
  & \raisebox{0pt}{$\Phi^*\!d_R\chi$ \ } & \multicolumn{2}{|c|}{\raisebox{1pt}{}}  \\[0pt]
\cline{1-4}
1C&  \raisebox{0pt}{$\!\!\!X_\mu Q_L \gamma^\mu d_R$ , $X^\dagger_\mu Q_L \gamma^\mu \chi$\!\!\!} & \multicolumn{2}{|c|}{$X_\mu = (3,2)_{\frac16}$ \ \ }   \\[2pt]
\hline
\hline
\multicolumn{4}{|c|}{\raisebox{0pt}{${\boldsymbol{qqq\chi\phi}}$}}\\[2pt]
\hline
\raisebox{-1pt}{2A} &  \raisebox{-1pt}{$\!\!\!\Phi \,d_R d_R$ , $\Phi^*\!u_R\widetilde\chi$ , $\overline{\widetilde\chi}\chi\phi$\!\!\!} & \raisebox{-10pt}{$\Phi = (3,1)_{\frac23}$} & \raisebox{-1pt}{$\widetilde\chi = (1,1)_0$}   \\[-8pt]
\cline{1-2}\cline{4-4}
\raisebox{-1.5pt}{2B} &  \raisebox{-1.5pt}{$\!\!\!\Phi\, d_R d_R$ , $\Phi^*\widetilde\chi\chi$ , $\overline{\widetilde\chi}u_R\phi$\!\!\!} &  &  \raisebox{-1pt}{$\widetilde\chi = (3,1)_{\frac23}$}   \\[3pt]
\cline{1-4}
\raisebox{-7pt}{2C}  & \raisebox{0pt}{\!$\big(\Phi \,u_Rd_R$    {\rm or} \,$\Phi \,Q_LQ_L$\big),} & \raisebox{-22pt}{$\!\Phi = (3,1)_{-\frac13}\!\!\!$} &  \raisebox{-6pt}{$\widetilde\chi = (1,1)_0$}  \\[-25pt]
  &\raisebox{1pt}{$\Phi^*\!d_R\widetilde\chi$ , $\overline{\widetilde\chi}\chi\phi \ \ $} & &  \\[-1pt]
\cline{1-2}\cline{4-4}
\raisebox{-7pt}{2D}  & \raisebox{0pt}{\!$\big(\Phi \,u_Rd_R$    {\rm or} \,$\Phi \,Q_LQ_L$\big),} &  & \raisebox{-6pt}{$\!\!\widetilde\chi = (3,1)_{-\frac13}\!\!\!\!$} \\[-9pt]
 & $\Phi^*\widetilde\chi\chi$ , $\overline{\widetilde\chi}d_R\phi$ \ \  \ &  &\\[1pt]
\cline{1-4}
\raisebox{-7pt}{2E}&  \raisebox{0pt}{$\!\!\!X_\mu Q_L\gamma^\mu d_R$ , $\overline{\widetilde\chi}\chi\phi$ ,\!\!\!} & \raisebox{-22pt}{$\!\!X_\mu = (3,2)_{\frac16}\!\!\!\!$} & \raisebox{-7pt}{$\widetilde\chi = (1,1)_0$}   \\[-25pt]
&  \raisebox{0pt}{$X^\dagger_\mu Q_L \gamma^\mu\widetilde\chi\!\!\!$} &  &   \\[1pt]
\cline{1-2}\cline{4-4}
\raisebox{-7pt}{2F}&  \raisebox{0pt}{$\!\!\!X_\mu Q_L \gamma^\mu d_R$ , $X^\dagger \widetilde\chi\chi$ ,} & & \raisebox{-7pt}{$\widetilde\chi = (3,2)_{\frac16}$}   \\[-10pt]
&  \raisebox{0pt}{ $\overline{\widetilde\chi}Q_L\phi$} & &   \\[1pt]
\hline
\end{tabular}
\endgroup
\end{center}
\vspace{-2mm}
\caption{Effective operators describing the interaction of baryonic dark matter with quarks and their model realizations.}\vspace{6mm}
\label{tab33B}
\end{table}

\vspace{6mm}

\noindent
{\bf{Model 1A \  ($\boldsymbol{qqq\chi}$)}}
\vspace{2mm}

\noindent
The Lagrangian for Model 1A is given by
\bea\label{llag1}
- \mathcal{L}_1 &\supset&  \lambda_q^{ab} \epsilon^{ijk} \Phi_i d_{Rj}^a d_{Rk}^b  + \lambda_\chi^a \Phi^{*i} {\chi}u_{Ri}^a  \ + \ {\rm h.c.} \ , \ \ \ \ \ 
\eea
where $i,j,k$ are color indices and  $a,b$ are flavor indices. Due to the antisymmetric nature of the $\epsilon$ tensor, the coupling
$\lambda_q^{ab}$ must be antisymmetric in flavor.

Although there is no tree-level proton decay to a final state consisting of only Standard Model particles, the model still suffers from proton decay if $m_\chi < m_p - m_e = 937.761 \ {\rm MeV}$, since the proton can then undergo the dark decay $p \to \bar\chi\,e^+ \nu_e$. 
The mass range $937.761 \ {\rm MeV} < m_\chi < 937.993 \ {\rm MeV}$ is  problematic  as well -- although the proton remains stable, the $^9{\rm Be}$ nucleus, known to be stable, can then undergo the nuclear dark decay $^9{\rm Be} \to \chi + 2\,\alpha$ \cite{Pfutzner:2018ieu}.

 The mass range 
$
937.993 \ {\rm MeV} < m_\chi < 938.783 \ {\rm MeV} 
$
is of particular interest, since then the  proton,  all stable nuclei and the dark matter $\chi$ remain stable,  but the neutron can undergo the dark decay $n \to \bar\chi\,\gamma$. This decay channel was 
proposed in \cite{Fornal:2018eol} as a possible solution to the neutron lifetime discrepancy. It was shown that Model 1B allows the neutron to have a dark decay branching fraction ${\rm Br}(n\to \bar\chi\,\gamma)  = 1\%$, which corresponds to  $m_\Phi \sim \mathcal{O}(100 \ {\rm TeV})$ for order one couplings; see \cite{Fornal:2020gto} for a 
 detailed review of this proposal along with its experimental signatures. It was recently argued  that within the framework of Model 1A, the neutron dark decay branching fraction is constrained to be ${\rm Br}(n\to \bar\chi\,\gamma) < 10^{-6}$ \cite{Fajfer:2020tqf}.
 
In Models 1A--1C, the dark matter $\chi$ can annihilate with the neutron, leading to  signatures such as $\chi \,n \to \gamma + {\rm meson(s)}$ at Super-Kamiokande  and the future DUNE \cite{Keung:2019wpw}. We will discuss this in more detail in the context of Model 2A below.
Finally, in the case $m_\chi > m_p+ m_e = 938.783 \ {\rm MeV}$ the dark particle $\chi$ is unstable, since the decay channel $\chi \to \bar{p}\, e^+\nu_e$ opens up kinematically. Although in this scenario $\chi$ is not the dark matter, it can still be produced in experiments and lead to detectable signatures in $B$ factories, e.g.,  missing energy signals from dark decays of heavy baryons and mesons \cite{Heeck:2020nbq}.

\vspace{5mm}

\noindent
{\bf{Model 2A \  ($\boldsymbol{qqq\chi\phi}$)}}
\vspace{2mm}

\noindent
The Lagrangian for Model 2A is given by a simple extension of the interactions in Eq.\,(\ref{llag1}),
\bea\label{llag2}
-\mathcal{L}_2 \ &\supset& \  \lambda_q^{ab} \epsilon^{ijk} \Phi_i d_{Rj}^a d_{Rk}^b  + \lambda_{\widetilde\chi}^a \Phi^{*i} {\widetilde\chi}u_{Ri}^a \nn\\
&& + \   \lambda_\phi\overline{\widetilde\chi}\chi\phi  \,+ {\rm h.c.} \ . 
\eea
Once again, the coupling $\lambda_q^{ab}$ is antisymmetric in flavor.
The stability of the proton and all stable nuclei is guaranteed by the condition $m_\chi + m_\phi > 937.993 \ {\rm MeV}$. In the special case   $937.993 \ {\rm MeV} < m_\chi + m_\phi < m_n = 939.565 \ {\rm MeV}$, the neutron can undergo the decay $n \to \bar\chi\,\phi^*$. This is the second neutron dark decay channel proposed in \cite{Fornal:2018eol} to solve the neutron lifetime discrepancy, and leads to  unique signals in nuclear decays \cite{Pfutzner:2018ieu}. 
In contrast to Model 1A, a large mass of $\chi$ or $\phi$ does not necessarily lead to dark matter decay;\break provided that $|m_\chi - m_\phi| < m_p+m_e = 938.783 \ {\rm MeV}$, both $\chi$ and $\phi$ remain stable.

Upon including an additional heavy particle $\widetilde\chi'$  with the coupling $\overline{\widetilde\chi'}\chi\phi$, non-trivial $CP$ phases can lead to different decay probabilities for $\widetilde\chi \to udd$ versus $\overline{\widetilde\chi} \to \bar{u}\bar{d}\bar{d}$ through the interference between tree-level and loop-level decay channels.
Model 2A is then capable of explaining the matter-antimatter asymmetry of the universe through the   hylogenesis mechanism \cite{Davoudiasl:2010am} if $m_\chi+m_\phi \sim  5 \ {\rm GeV}$.

This model also  predicts striking signatures in direct detection experiments:  $\chi \, N \to \phi^* + {\rm meson(s)}$, where $N$ is a proton or a neutron. 
Such signals were  studied in  \cite{Davoudiasl:2011fj,Demidov:2015bea,Jin:2018moh,Keung:2019wpw} and include: $\chi \, n \to \phi^* \, \pi^0$, \   $\chi \, p \to \phi^* \, \pi^+$,  \ $\chi \, n \to \phi^* \, K^0$ and  \  $\chi \, p \to \phi^* \, K^+$. 
Interestingly, not all of them are present in Model 2A. Given the antisymmetric structure of $\lambda_q^{ab}$ in Eq.\,(\ref{llag2}),  the scalar mediator does not couple to two down quarks, which implies that signatures involving solely  pions in the final state are not possible in Model 2A.

In conclusion, we find that there is a specific prediction of baryonic dark matter models without tree-level proton decay: dark matter-nucleon annihilation  leads to at least one  kaon in the final state. This is an especially  relevant observation in light of DUNE's expected exquisite sensitivity to kaons. In certain regions of parameter space, DUNE will be able to probe $\Phi$ masses  up to $m_\Phi \sim \mathcal{O}(100 \ {\rm TeV})$ \cite{Keung:2019wpw}.

\vspace{7mm}

\noindent
\begin{centerline}
{\bf{Baryoleptonic dark matter}}
\end{centerline}

The simplest class of baryoleptonic dark matter models arises from dimension seven effective operators $qqql\phi$ and $qqq\bar{l}\phi$, where $l$ is a Standard Model lepton representation. Minimal particle physics realizations of  $qqql\phi$
involve one of the scalar mediators $(3,1)_{-1/3}$, $(3,1)_{-4/3}$, or one of the vectors $(3,2)_{-5/6}$, $(3,2)_{1/6}$, whereas particle models for  $qqq\bar{l}\phi$ require the scalar mediator $(3,1)_{-1/3}$ or the vector $(3,2)_{1/6}$. 
In models with only a single  mediator,  it has to couple to the bilinears $qq$ and $ql$ (or $q\bar{l}$),  resulting in tree-level proton decay which cannot be forbidden by any symmetry.

A possible way to overcome this issue is to  introduce two mediators, one coupled only to $qq$, and  the other coupled only to $ql$ (or $q\bar{l}$) \cite{Heeck:2020nbq}, or include an additional heavy fermion \cite{Huang:2013xfa}. However, we will not consider those scenarios here, since they require imposing baryon/lepton number conservation.  A simple extension of this class of models,  described  by the effective operator  $qqql\phi^2$, i.e., replacing $\phi$ by  $\phi_e$  and adding the interaction term $\phi_e^* \phi^2$ to stabilize  $\phi$, provides a working mechanism for baryogenesis as in Model 2A and also exhibits nonstandard nucleon destruction signatures \cite{Huang:2013xfa}. A more general analysis of the baryon asymmetry generation in those types of models was performed in \cite{Bernal:2016gfn}.

\vspace{7mm}

\noindent
\begin{centerline}
{\bf{Leptonic dark matter}}
\end{centerline}

In order to explain the matter-antimatter asymmetry through leptogenesis, operators carrying a nonzero lepton number contribution from the leptons are needed. Those operators can generate an asymmetry in the lepton sector, which is then transferred to the baryon sector by electroweak sphalerons \cite{Klinkhamer:1984di}. 
The dimension four and five operators $l\bar{l} \phi$ and $l\bar{l} \phi^2$ are not of interest to us since $\phi$ does not  carry lepton number.  Also, as in the baryonic dark matter case, we will not consider operators involving the Higgs, e.g., the dimension four operator $H L_L\chi$.

\vspace{5mm}

\begin{figure}[t!]
\includegraphics[trim={0.0cm 0.0cm 0 0},clip,width=4.8cm]{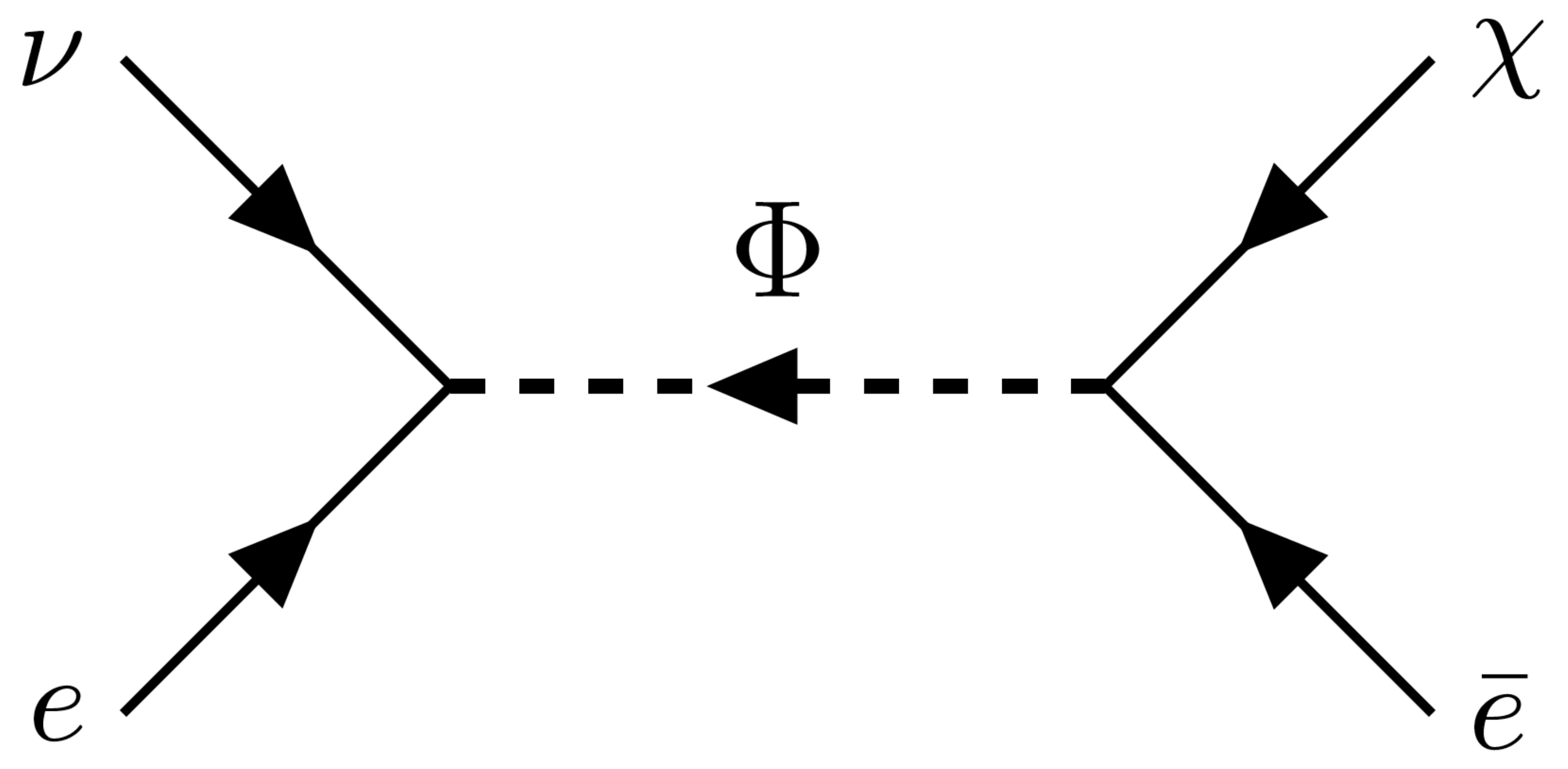}
\caption{{\small{Model 3A realization of the operator $ll\bar{l}\chi$.} \vspace{4mm} }}
\label{fig:3}
\end{figure}

The first operator of interest is the dimension six $ll\bar{l}\chi$, where  $l$ can be either $L_L$ or $e_R$. The only gauge-invariant realization is $L_LL_L\overbar{e_R}\chi$. Note that the operator $lll\chi$ is not  invariant under hypercharge. 
There are two particle models  for the operator $ll\bar{l}\chi$, denoted as Models 3A and 3B in Table \ref{tab44L}. 
The corresponding mediators are the scalars $(1,1)_{1}$ and $(1,2)_{-1/2}$, respectively.
They are colorless, thus they do not mediate proton decay. 
Because of the similarity between the two models, we write down the Lagrangian only for Model 3A. Its realization of the operator $ll\bar{l}\chi$  is shown in Fig.\,\ref{fig:3}.

\begin{table}[t!] 
\begin{center}
\begingroup
\setlength{\tabcolsep}{6pt} 
\renewcommand{\arraystretch}{1.5} 
\begin{tabular}{  |c |c |c|c|} 
\hline
 \multicolumn{4}{|c|}{Leptonic dark matter}\\
\hline
\!\!Model\!\! & Interactions &  \multicolumn{2}{|c|}{Mediators} \\ 
\hline
\hline
\multicolumn{4}{|c|}{\raisebox{-1pt}{${\boldsymbol{ll\bar{l}\chi}}$}}\\[2pt]
\hline
3A &  \raisebox{0pt}{$\Phi L_LL_L$ , $\Phi^*\overbar{e_R}\chi$} & \multicolumn{2}{|c|}{$\Phi = (1,1)_{1}$}  \\[2pt]
\cline{1-4}
\raisebox{0pt}{3B}  & \raisebox{0pt}{$\Phi L_L\overbar{e_R}$ , $\Phi^* \!{L_L}\chi$} & \multicolumn{2}{|c|}{\raisebox{1pt}{$\Phi = (1,2)_{-\frac12}$}}  \\[1pt]
\hline
\hline
\multicolumn{4}{|c|}{\raisebox{-1pt}{${\boldsymbol{ll\bar{l}\chi\phi}}$}}\\[2pt]
\hline
4A &  \raisebox{0pt}{\!$\Phi L_LL_L$ , $\Phi^*\overbar{e_R}\widetilde\chi$ , $\overline{\widetilde\chi}\chi\phi$\!} & \raisebox{-8pt}{$\Phi = (1,1)_{1}$} & $\widetilde\chi = (1,1)_0$   \\[-5pt]
\cline{1-2}\cline{4-4}
4B &  \raisebox{0pt}{\!$\Phi L_LL_L$ , $\Phi^*\widetilde\chi\chi$ , $\overline{\widetilde\chi}\overbar{e_R}\phi$\!} &  &  $\widetilde\chi = (1,1)_{1}$   \\[2pt]
\cline{1-4}
4C&  \raisebox{0pt}{\!$\Phi L_L \overbar{e_R}$ , $\Phi^* L_L\widetilde\chi$ , $\overline{\widetilde\chi}\chi\phi$\!} & \raisebox{-7pt}{$\!\Phi = (1,2)_{-\frac12}\!\!\!$} & $\widetilde\chi = (1,1)_0$   \\[-6pt]
\cline{1-2}\cline{4-4}
4D&  \raisebox{0pt}{\!$\Phi L_L \overbar{e_R}$ , $\Phi^* \widetilde\chi\chi$ , $\overline{\widetilde\chi}L_L\phi$\!} & & $\widetilde\chi = (1,2)_{-\frac12}$ \!\! \!\!\!\! \\[2pt]
\hline
\end{tabular}
\endgroup
\end{center}
\vspace{-2mm}
\caption{Effective operators describing the interaction of leptonic dark matter with leptons and their model realizations.}\vspace{5mm}
\label{tab44L}
\end{table}

\begin{figure}[t!]
\includegraphics[trim={0.0cm 0.0cm 0 -0.4cm},clip,width=5.7cm]{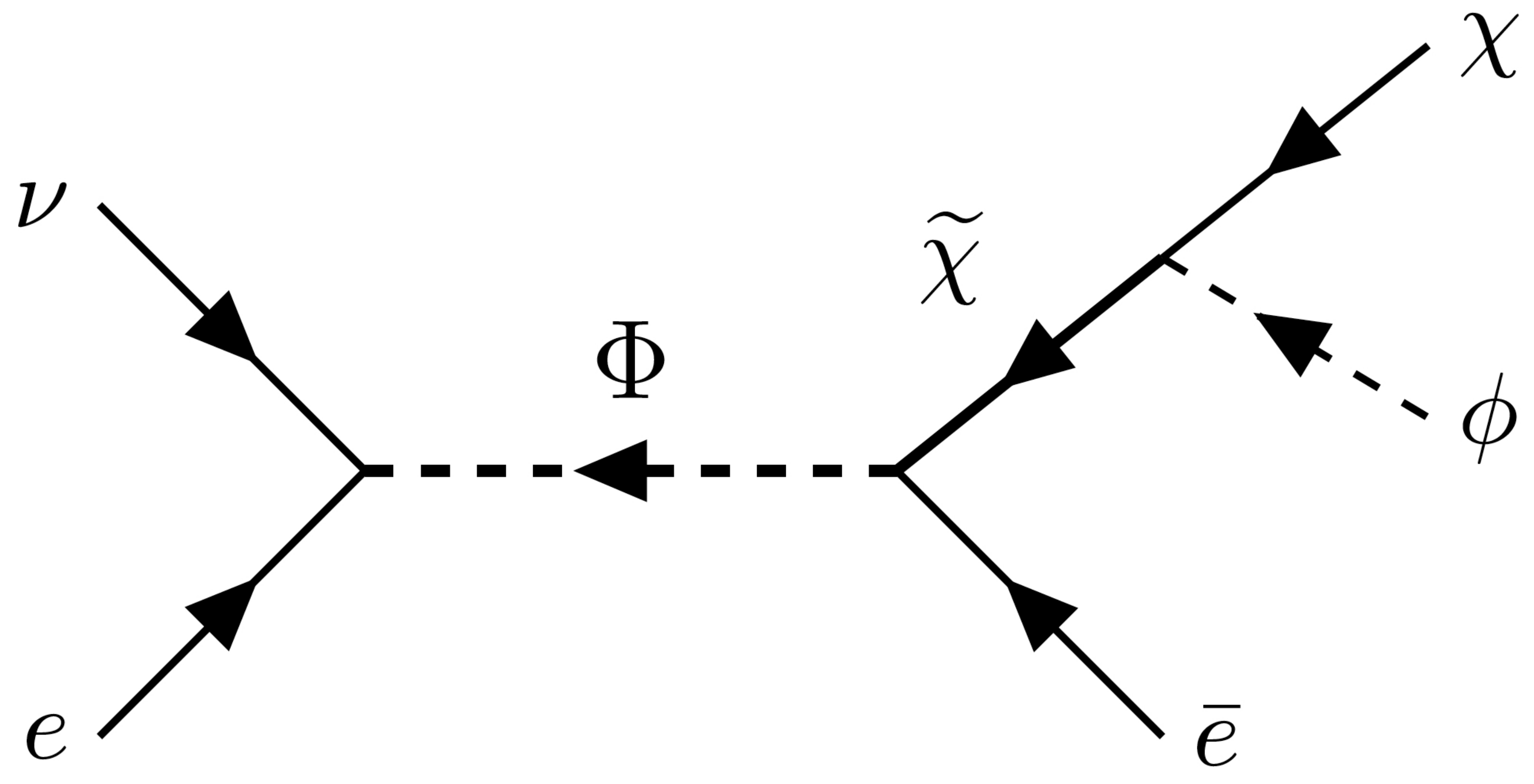}
\caption{{\small{Model 4A realization of the operator $ll\bar{l}\chi\phi$.}  \vspace{4mm} }}
\label{fig:4}
\end{figure}

At dimension seven, the effective operator of interest is  $ll\bar{l}\chi\phi$, again with only a single   gauge-invariant realization  $L_LL_L\overbar{e_R}\chi\phi$. 
The corresponding particle models are labeled as Models 4A--4D in Table \ref{tab44L}. An intermediate particle $\widetilde\chi$ is required. 
We focus on Model 4A below for a  quantitative discussion of its properties. 
The realization of  the operator $ll\bar{l}\chi\phi$ within this model is presented  in Fig.\,\ref{fig:4}.

Regarding  other options, not involving the Higgs field, the dimension seven operator $ll\bar{l}\bar{l}\phi$ does not carry nonzero lepton number, thus it cannot fit into the framework of asymmetric dark matter. At dimension eight, the only operator carrying  lepton charge is $ll\bar{l}\chi\phi^2$. As in the  baryonic dark matter case, models for this operator are obtained from Models 4A--4D by substituting $\phi$ with $\phi_e$, and introducing the interaction $\phi_e^*\phi^2$. This makes  $\phi$ automatically stable, but offers no other advantages  compared to the $ll\bar{l}\chi\phi$ case. 

\vspace{5mm}

\noindent
{\bf{Model 3A \  ($\boldsymbol{ll\bar{l}\chi}$)}}
\vspace{2mm}

\noindent
The Lagrangian for Model 3A is
\bea\label{eee3}
- \mathcal{L}_3 \ &\supset& \   \lambda_l^{ab} \Phi(L_{L}^a \epsilon L_{L}^b)  + \lambda_{\chi}^{a} \,\Phi^{*}\! {\chi}\overbar{e_R}^a  +  {\rm h.c.} \ ,\ \ \ \ \ 
\eea
where again   $a,b$ are flavor indices and the parenthesis denotes the contraction of ${\rm SU}(2)_L$ indices. 
Because of the antisymmetric structure of the $LL$ bilinear, the coupling $\lambda_l^{ab}$ must be antisymmetric in flavor. 

This model is not phenomenologically attractive because it is hard to ensure the stability of the dark matter particle. Especially, without imposing any lepton flavor symmetry, $\chi$ can undergo the decay $\chi \to \nu \gamma$ through a loop diagram. Thus, we do not consider this class of models further.

\vspace{6mm}

\noindent
{\bf{Model 4A \  ($\boldsymbol{ll\bar{l}\chi\phi}$)}}
\vspace{2mm}

\noindent
A phenomenologically viable model which contains  a dark matter candidate and  can successfully explain the matter-antimatter of the universe is obtained by introducing an additional  scalar particle $\phi$.
The resulting Lagrangian is a minimal extension of $\mathcal{L}_3$,
\bea\label{llag4}
- \mathcal{L}_4 \ &\supset& \   \lambda_l^{ab}  \Phi (L_{L}^a \epsilon L_{L}^b)  + \lambda_{\widetilde\chi}^{a} \,\Phi^{*} {\widetilde\chi}\overbar{e_R}^a  + \lambda_\phi \overline{\widetilde\chi}\chi\phi +  {\rm h.c.} \ . \ \ \ \ \ 
\eea
The same symmetry arguments as for Model 3A apply, i.e., $  \lambda_l^{ab}$ is antisymmetric in flavor.  In addition, we imposed a $Z_2$ parity in the dark sector, under which $\phi$ and $\chi$ are odd, so that the lighter of them remains stable and can constitute the dark matter. Model 4A is the subject of the subsequent section.

\vspace{4mm}

\section{Phenomenology of leptonic dark matter}\label{s4}

The crucial property of leptonic dark matter models is that they 
 do not suffer from proton decay.  Below we discuss the baryon asymmetry generation in Model 4A via leptogenesis, the symmetry restoration in the dark sector  and the signatures expected in indirect detection experiments. Our conclusions apply  to Models 4B--4D as well.

\vspace{6mm}

\noindent
\begin{centerline}
{\bf{Leptogenesis}}
\end{centerline}

The generation of lepton asymmetry within the framework of Model 4A is similar to the hylogenesis mechanism \cite{Davoudiasl:2010am}.
The process
starts  immediately  after inflation. Once the inflaton field $\Psi$  falls into the potential minimum, it starts oscillating and leads to the reheating of the universe. We assume that during reheating the particles $\widetilde\chi$ and $\overline{\widetilde\chi}$ are produced in equal amounts. 
As the temperature drops, the particles $\widetilde\chi$ decay via the channels $\widetilde\chi \to \chi\,\phi$ and $\widetilde\chi \to e_R \overbar{e_L} \overbar{\nu_L}$, where $\overbar{e_L} $ and $\overbar{\nu_L} $ have different flavors. The antiparticles $\overline{\widetilde\chi}$ decay through conjugate processes.

\begin{figure}[t!]
\includegraphics[trim={0.0cm 0.0cm 0 0},clip,width=8.4cm]{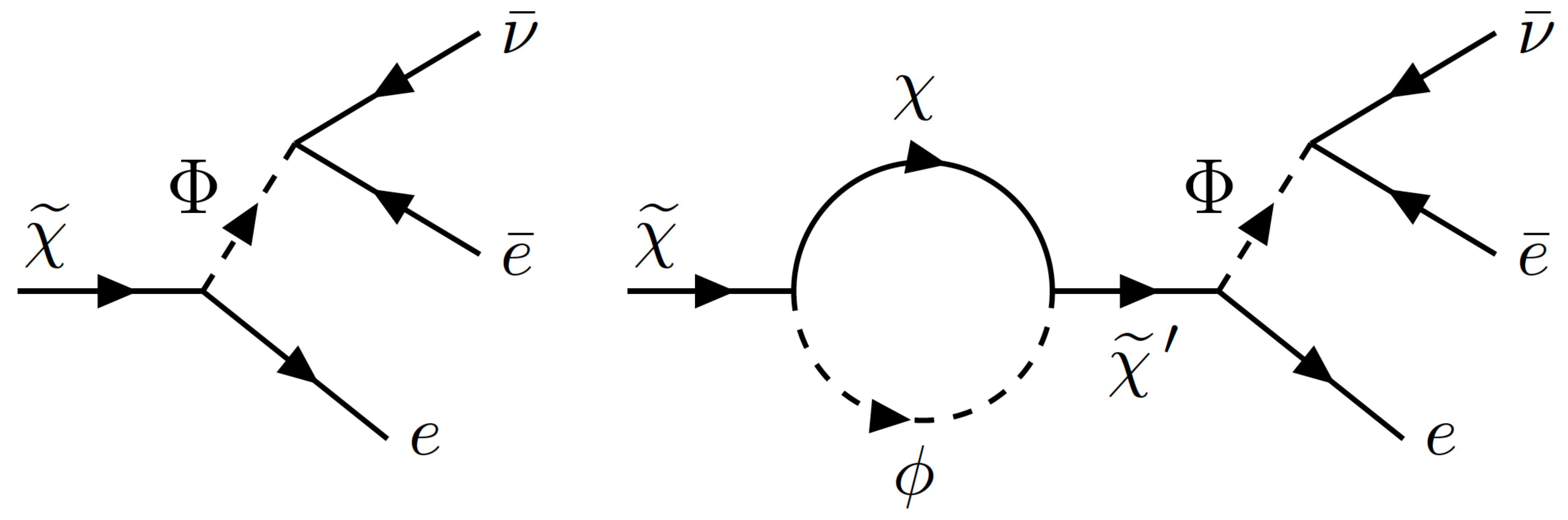}
\caption{Diagrams contributing to the decay $\widetilde\chi \to e_R \overbar{e_L} \overbar{\nu_L}$.\vspace{2mm}}
\label{fig:5}
\end{figure}

In order to generate $CP$ violation in the model, we introduce the particle $\widetilde\chi'$ with the following interaction  terms,
\bea
-  \mathcal{L}'_4 \ \supset \  \lambda_{\widetilde\chi'}^{a} \,\Phi^{*} {\widetilde\chi'}\overbar{e_R}^a  + \lambda'_\phi\, \overline{\widetilde\chi'}\chi\phi +  {\rm h.c.}  \ ,
 \eea
and we take $m_{\widetilde\chi'} \gg m_{\widetilde\chi}$. 
The asymmetry between the decays of ${\widetilde\chi}$ and $\overline{\widetilde\chi}$ arises from the interference between the tree-level and one-loop diagram shown in Fig.\,\ref{fig:5}. 
If  the leading decay channel is  $\widetilde\chi \to \chi\,\phi$, 
the generated lepton asymmetry is \footnote{In this calculation, we only consider the decay channel with one particular lepton flavor choice. The rescaling can be easily done to include more flavor channels.}
\bea
\Delta_L \ &=& \ \frac{\Gamma(\widetilde\chi \to e_R\overbar{e_L}\overbar{\nu_L})-\Gamma(\overline{\widetilde\chi} \to \overbar{e_R}{e}_L \nu_L)}{2\,\Gamma(\widetilde\chi \to \chi\,\phi)}\nn\\[6pt]
&\approx& \ \frac{|\lambda_l|^2\,{\rm Im} \big(\lambda_{\widetilde\chi}^{*}\lambda_{\widetilde\chi'}\lambda_{\phi}\lambda_{\phi}^{\prime*}\big)}{{1536}\,\pi^3 |\lambda_\phi|^2 } \frac{m_{\widetilde\chi}^5}{m_\Phi^4 \,m_{\widetilde\chi'}}  \ . \ \ \ 
\eea
 In order to avoid the washout of the  asymmetry by $\chi\,\phi \to e_R \overbar{e_L}  \overbar{\nu_L}$ scattering, the reheating temperature, for couplings ${\mathcal{O}(1)}$, needs to satisfy \cite{Davoudiasl:2010am}
\bea\label{reheat}
T_R \lesssim {(11 \ {\rm TeV})}\bigg[{\frac{m_\Phi^4  m_{\widetilde\chi}^2}{(1 \ {\rm PeV})^6}}\bigg]^{\!\frac15} \! .
\eea

Provided that $T_R$ is above the electroweak symmetry breaking scale, i.e., $T_R \gtrsim 200 \ {\rm GeV}$, a portion of the created  lepton asymmetry will be converted into a baryon asymmetry by the electroweak sphalerons. Therefore, as long as the masses of $\Phi$ and $\widetilde\chi$ fulfill the condition 
\bea\label{bbound}
(m_\Phi^4 m_{\widetilde\chi}^2)^{\frac16}\gtrsim {40 \ {\rm TeV}} \ ,
\eea
the sphalerons efficiently  transfer the asymmetry to the baryonic sector. 
We find that the final baryon asymmetry is
\bea
\Delta_B \approx \frac{28}{79}\,\Delta_L \ .
\eea
The exact relation between  the baryon-to-photon ratio $\eta_B$ and the baryon asymmetry depends on the model of reheating; up to $\mathcal{O}(1)$  factors it is given by \cite{Asaka:1999yd,Giudice:1999fb,Buchmuller:2005eh}
\bea
\eta_B \approx \frac{\Delta_B T_R}{M_\Psi} \ ,
\eea
where $M_\Psi$ is the  inflaton mass.  The observed value of the baryon-to-photon ratio in the universe of $\eta_B \approx 6 \times 10^{-10}$ \cite{Aghanim:2018eyx}  is obtained, e.g., for  $\mathcal{O}(1)$ couplings,  {$ m_\Phi = m_{\widetilde\chi} = 1 \ {\rm PeV}$, \,$m_{\widetilde\chi'} = 5 \ {\rm PeV}$, \,$M_\Psi = 20 \ {\rm PeV}$\, and \, $T_R = 8  \ {\rm TeV}$.}

 The sum of $\chi$ and $\phi$ masses is set by the observed ratio of the dark matter and baryon abundances. Since the generated dark matter asymmetry $\Delta_{\rm DM} = \Delta_L$, one arrives at
 \bea\label{aadm}
 m_\chi + m_\phi = m_p \frac{\Omega_{\rm DM}}{\Omega_B} \,\bigg| \frac{\Delta_B}{\Delta_{\rm DM}}\bigg|  \approx 1.8 \ {\rm GeV} \ .
 \eea

\begin{figure}[t!]
\includegraphics[trim={0.0cm -0.8cm 0 0},clip,width=8.0cm]{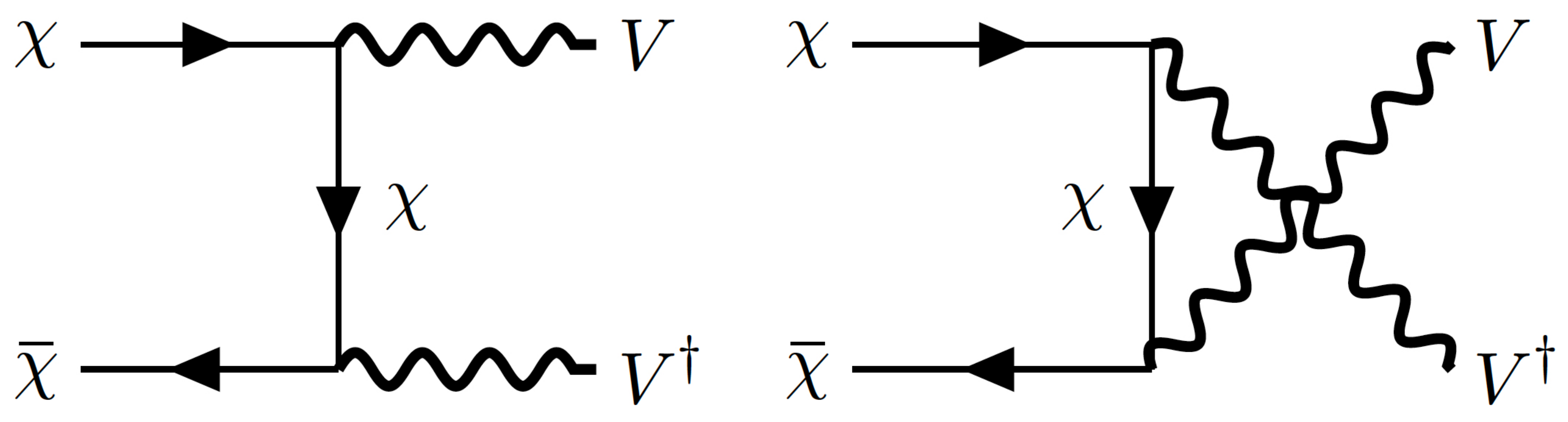}
\caption{Annihilation of the dark matter symmetric component.\vspace{2mm}}
\label{fig:ann}
\end{figure}

\vspace{6mm}

\noindent
\begin{centerline}
{\bf{Removing the symmetric component}}
\end{centerline}

To ensure the annihilation of the  symmetric component of dark matter, it is sufficient to introduce one new particle, e.g., a vector  $V$, lighter than $\chi$ and $\phi$, with the interaction terms
\bea\label{annn}
- \mathcal{L}''_{4} \  &\supset& \ g_{\chi}V_\mu\,\overbar\chi\gamma^\mu \chi + g_{\phi}  V_\mu\,\phi\, \partial^\mu \phi^* \nn\\
&& + \    g_{V}  V_\mu\, \overbar{e_R}^a \gamma^\mu e_R^a   + {\rm h.c.}\ . 
\eea 
This  allows for the annihilation channels $\overbar\chi \chi \to V V^\dagger$ and  $\phi\, \phi^* \to VV^\dagger$, along with the subsequent decays $V \to l^+ l^-$. 
Diagrams corresponding to $\overbar\chi \chi \to V V^\dagger$ are shown in Fig.\,\ref{fig:ann}. The annihilation cross section $\sigma_{\overbar\chi \chi}$ is related to the symmetric component of the $\chi$ relic density as
\bea\label{Omegas}
\Omega_\chi h^2  \simeq \left(\frac{9\times 10^{-11}}{{\rm GeV^2}}\right)\frac{1}{\langle\sigma_{\overbar\chi\chi}v\rangle \sqrt{g_*}}\frac{m_\chi}{T_f} \ ,
\eea
where $g_*$ is the number of relativistic degrees of freedom and $T_f$ is the freeze-out temperature \cite{Kolb:1990vq}.
Given the observed value of the dark matter relic density $\Omega_{\rm DM}h^2 = 0.12$ \cite{Tanabashi:2018oca},  we find that for $m_V \ll m_\chi$  an efficient annihilation of the symmetric component of $\chi$ is achieved when
\bea
g_\chi \gtrsim 0.01 \ . 
\eea
Similar arguments  apply to the annihilation  $\phi\, \phi^* \to V V^\dagger$.  The decays of $V$, such as $V \to e^+e^-$, depending on its coupling and mass, can be a slow process. Thus  a small coupling between $V$ and leptons is allowed, and can easily be consistent with all experimental constraints \cite{Jeong:2015bbi}.

Another possibility is to introduce a light scalar $\phi'$, rather than the vector boson $V$, to remove the symmetric components of $\chi$ and $\phi$. The annihilation $\overbar\chi \chi \to \phi' \phi^{\prime *}$ is a $p$-wave process, thus it is suppressed by the relative velocity. Such a velocity suppression does not significantly affect the annihilation of the symmetric components, since both $\chi$ and $\phi$ are still semi-relativistic during their freeze-out. However, it has interesting implications for the experimental constraints and signatures, which will be discussed below.

\vspace{5mm}

\noindent
\begin{centerline}
{\bf{Symmetry restoration}}
\end{centerline}
\vspace{-1mm}

\begin{figure}[t!]
\includegraphics[trim={0.0cm 0.3cm 0 -0.0cm},clip,width=5.4cm]{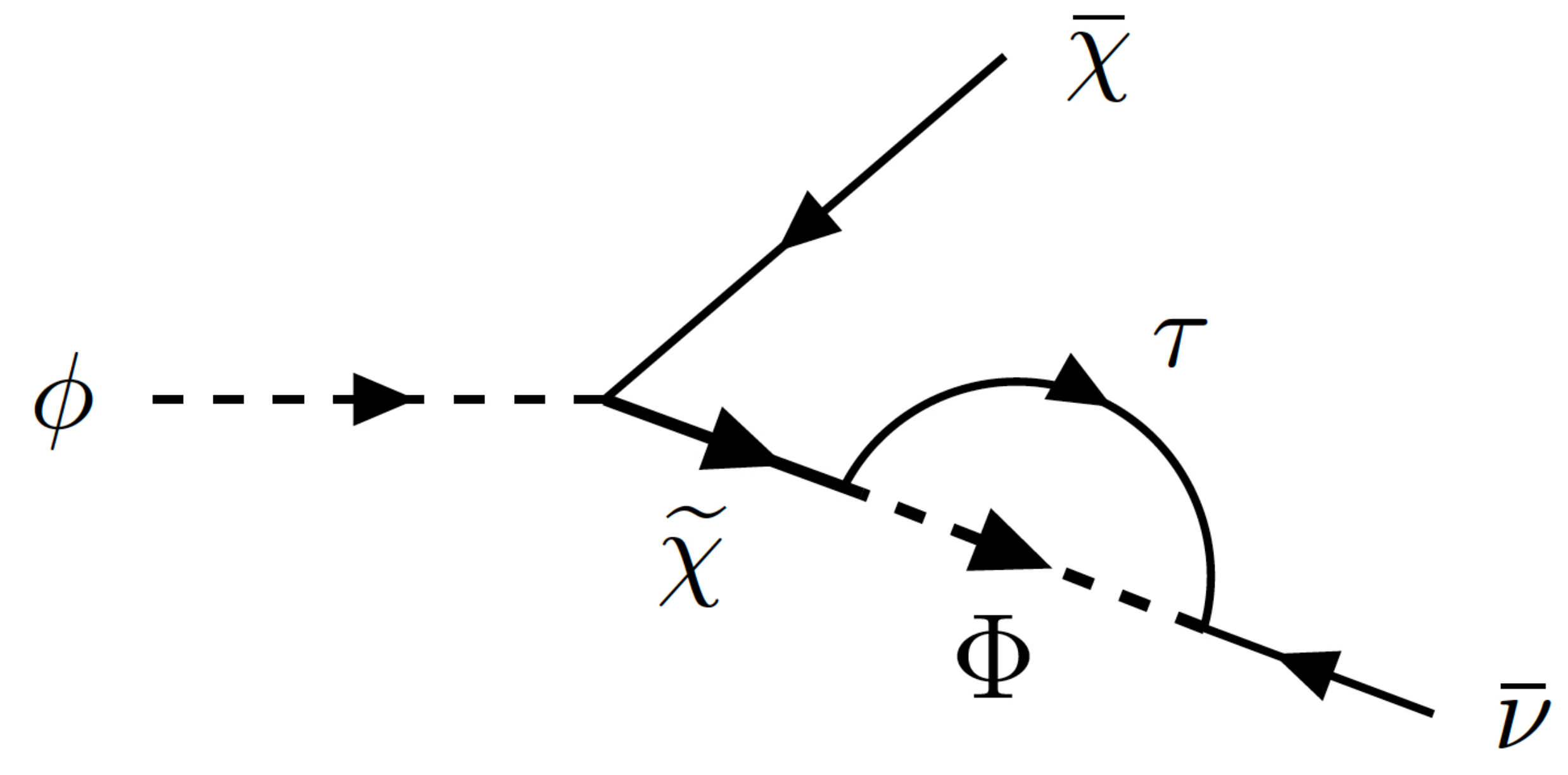}
\caption{{\small{Diagram for the decay $\phi \to \overbar\chi\,\overbar\nu$.}  \vspace{0mm} }}
\label{fig:6}
\end{figure}

Interestingly, within Model 4A the  asymmetry in the dark sector is not preserved  during  the evolution of the universe. 
If  $m_\phi > m_\chi$,  the interactions in Eq.\,(\ref{llag4}) render the particle $\phi$ unstable, resulting in the  two-body decay $\phi \to \overbar\chi\,\overbar\nu$ and the four-body decay $\phi \to \overbar\chi\,\overbar\nu\,e^+e^-$. The two-body channel is  dominant and proceeds through the diagram shown in Fig.\,\ref{fig:6}.  
The resulting decay  rate  is 
 \bea
\Gamma(\phi \to \overbar\chi \,\bar\nu) &\sim&   \frac{|\lambda_l^{} \,\lambda_{\widetilde\chi} \lambda_\phi|^2}{4096\,\pi^5}  \,\frac{m_\tau^2\, m_\phi}{m_{\widetilde\chi}^2}  \bigg(1-\frac{m_\chi^2}{m_\phi^2}\bigg)^{\!2} \ .  \ \ \ 
\eea
The observed dark matter relic density imposes a constraint on this rate. More explicitly, the decays $\phi \to \overbar\chi\,\overbar\nu$ restore the symmetry between $\chi$ and $\bar \chi$ in the relic abundance. Such a restoration must happen sufficiently late so that $\chi$ and $\overbar\chi$ do not efficiently  annihilate with each other to cause $\mathcal{O}(1)$ change {to} the dark matter relic abundance. We find that this requirement is met if the decay happens {at temperatures} $T\lesssim 50$ MeV or, {equivalently, after} $t\simeq O(10^{-4})$ s. This can be achieved if the mass splitting between $\chi$ and $\phi$ is small. For example, {taking} $\mathcal{O}(1)$ couplings, $m_{\widetilde\chi} \sim 1 \ {\rm PeV}$ and $(m_\phi-m_\chi)/m_\phi \sim 10^{-3}$, the  lifetime of $\phi$  {is $\tau_\phi \sim$} $0.1 \ {\rm s}$.
Since the decays of $\phi$'s only produce slow-moving $\chi$'s and low-energy neutrinos due to the small mass splitting, the cosmological constraints, e.g., from Big Bang Nucleosynthesis, are easily evaded \cite{Kanzaki:2007pd,Acharya:2020gfh}.

The restored symmetry in the dark sector revives   $\overbar\chi\chi$ annihilation at present times  in regions with large dark matter concentration, e.g., in the Galactic Center, which does not happen in standard asymmetric dark matter models.
  A similar scenario of symmetry restoration  in the dark sector was proposed  in  the context of  heavy dark matter \cite{Cohen:2009fz,Bell:2010qt,Falkowski:2011xh} and  oscillating dark matter \cite{Buckley:2011ye,Cirelli:2011ac,Tulin:2012re,Okada:2012rm,Ibe:2019yra}. In addition, late decays of the heavier dark matter component, $\tau_\phi \sim 10^9 \ {\rm years}$,  along with  a small mass splitting between the components,  are  consistent with observation \cite{Peter:2010jy,Bell:2010fk}. Such a late decay is also proposed as a possible solution to the  missing satellites problem
\cite{Abdelqader:2008wa} and the core-cusp problem \cite{Cline:2020gon}.

\vspace{6mm}

\noindent
\begin{centerline}
{\bf{Indirect detection signatures}}
\end{centerline}

The symmetry restoration  in the {dark sector} induces nontrivial dark matter indirect detection signals. For example, if $V$ or $\phi'$ couples to leptons and if its mass is larger than $2m_\mu$, the dark matter annihilation leads {to} final states  involving four leptons, i.e., $e^+e^-e^+e^-$, $e^+e^-\mu^+\mu^-$ or $\mu^+\mu^-\mu^+\mu^-$. Those leptons can further undergo  inverse Compton scattering and bremsstrahlung, producing photons that can be measured in indirect detection experiments.

The gamma ray constraints from the Fermi Gamma-Ray Space Telescope for a GeV-scale dark matter annihilating to four leptons {were} studied in  \cite{Abazajian:2010sq}. Depending on the species of the final state leptons, as well as on the mass of the light mediator, the {dark matter} annihilation rate is constrained to be $\langle \sigma_{\overbar\chi\chi} v\rangle \lesssim 10^{-25}\,-\, 10^{-24} \ {\rm cm^3}/{\rm s}$. 

A more stringent constraint {arises} from the {measurements of the} Cosmic Microwave Background {(CMB)} provided by the Planck satellite \cite{Aghanim:2018eyx}. More explicitly, $\langle \sigma_{\overbar\chi\chi} v\rangle $ is required to be smaller than $ 10^{-27} \ {\rm cm^3}/{\rm s}$ \cite{Slatyer:2015jla}. However, such a constraint is evaded if the light mediator is a scalar, i.e.{,} $\phi'$, {since} the dark matter annihilation cross section is largely reduced due to the velocity suppression at {late} times. In addition, if $\phi$ or the light mediator ($V$ or $\phi'$) is very long-lived due to either a small mass splitting or tiny couplings, the energy deposition to the Standard Model thermal bath only happens after the recombination, and the CMB constraint is also avoided.

{Lastly}, dark matter annihilation to a four-lepton final state  will be probed independently by the future e-ASTROGAM experiment  in the region  $\langle \sigma_{\overbar\chi\chi} v\rangle \gtrsim 4\times 10^{-27} \ {\rm cm^3}/{\rm s}$ \cite{Bartels:2017dpb}.

\section{Summary}

The nature of dark matter, origin of the matter-antimatter asymmetry of the universe and proton stability  are certainly among the greatest open questions in modern particle physics.  Theories  in which dark matter couples to quarks and leptons introduce a natural framework for solving the first two of those puzzles. 
However, many of  such models suffer from tree-level proton decay unless a conservation of baryon and/or lepton number is imposed by hand.

In this paper, we chose  proton stability as the primary criterion for a model's naturalness. We demonstrated  that this reduces the number of viable baryonic dark matter theories  to just a few models involving the scalar mediator $(3,1)_{2/3}$. This has an interesting impact on the potential experimental signatures -- we found that for baryonic  models  naturally free from  proton decay, the final state of dark matter-nucleon annihilation necessarily  involves a kaon.

We also considered a model of  leptonic dark matter, which explains the matter-antimatter asymmetry of the universe through leptogenesis and predicts symmetry restoration in the dark sector. In this theory the dark matter annihilation {may} be enhanced at late times, providing signals that can be  searched for in future indirect detection experiments.

\subsection*{Acknowledgments} 

The authors are supported in part by the U.S. Department of Energy under Award
No.~${\rm DE}$-${\rm SC0009959}$.

\bibliography{DM}

\end{document}